\documentclass[prl,twocolumn,showpacs]{revtex4}
\usepackage{epsfig}
\usepackage{dcolumn}
\usepackage{epstopdf}
\usepackage{graphicx}
\usepackage{subfigure}
\usepackage{epsf}
\begin{document}
\title{Fermi-surface pockets in magnetic  underdoped cuprates from first principles
}
\author{Alessio Filippetti}
\author{Danilo Puggioni}
\author{Vincenzo Fiorentini}
\affiliation{CNR-INFM SLACS, and Dipartimento
di Fisica, Universit\`a di Cagliari, I-09042 Monserrato (CA), Italy}

\date{\today}
\begin{abstract}
Using an innovative first-principles band theory enabling the exploration of Mott-insulating magnetic cuprates, we study the Fermi surface of underdoped Y$_{1-x}$Ca$_x$Ba$_2$Cu$_3$O$_6$ in a selection of magnetically ordered and polaronic states.  All phases exhibit qualitatively similar, hole-like nodal-point small pockets. Their  properties (area, masses, mass sign) only partially match those extracted from recent quantum-oscillation experiments. Ab initio  calculations, therefore, do not  straightforwardly support a magnetic origin of quantum oscillations.
\end{abstract}

\maketitle

The existence of a Fermi surface (FS) in underdoped high-T$_c$ superconductors is attested  \cite{chak} by  angle-resolved photoemission 
(ARPES)  \cite{norman,hossain} and by recent Shubnikov-de Haas (SdH) and de Haas-van Alphen (dHvA) quantum oscillation 
observed in mildly underdoped ($\sim$0.1 holes per CuO$_2$ unit) YBa$_2$Cu$_3$O$_{6.5}$ and YBa$_2$Cu$_4$O$_{8}$ \cite{doiron,yelland124}. 
However, the FS detailed shape and origin continue to escape an unambiguous identification. Not only does SdH/dHvA indicate the existence of small Fermi pockets, whereas ARPES observes large zone-corner-centered cylinders at optimal doping turning into disconnected arcs near nodal points upon underdoping: in addition, the SdH/dHvA  Hall resistance sign suggests electron-like pockets, 
whereas all ARPES measurements consistently report hole-like arcs.

Reconciling arcs, pockets, and their electron vs hole nature is highly non trivial \cite{nota}. Ab initio calculations  \cite{noi,yelland124} on
the non-magnetic metal phase (the zero-order approximation to the normal state) of YBa$_2$Cu$_3$O$_{6.5}$ and YBa$_2$Cu$_4$O$_{8}$ indicate no solid evidence of FS pockets. Pockets may originate from a symmetry-breaking FS reconstruction, as suggested for ``1/8" compounds  \cite{millis-norman07}. Several  models of symmetry breaking  have appeared   recently, among which $d$-density-wave 
order \cite{chak}, field-induced long-range magnetic order \cite{zhang-field}, short-range magnetic order \cite{harrison,civelli,russi}, and 
magnetic polarons \cite{lozoi,keller}, but none of them has yet gained a general consensus. Magnetic correlations are popular  candidates as antiferromagnetic (AF) correlations coexist with, or survive into, the superconducting (SC) phase over a wide doping range, according to various  experimental probes  \cite{niedermayer,sanna}.

The very existence of a Fermi surface --hence of quasiparticles-- and the hints about the role of magnetism suggest a considerable scope 
for ab initio band-like quasiparticle calculations, accounting accurately for material-specific information, and for coupling to the lattice. A first-principles description of 
underdoped cu\-pra\-tes, however,  is very challenging. Standard density functional theory \cite{pickett} fails to describe the magnetic insulating Mott 
phase and predicts a non-magnetic metallic state at all dopings, in strident conflict with the observed magnetic correlations. Here we overcome this limitation using the  pseudo-self-interaction correction (pSIC) technique \cite{fs} (already applied satisfactorily to many correlated materials, including cuprates \cite{ff1,ff2,ff4}) to calculate the FS of the AF and magnetic polaron states in an underdoped cuprate. All the  states we investigate  exhibit nodal pockets. In some cases, the pocket area is not far from experiment; the value and sign of the cyclotron masses, however, prevent  a conclusive identification, leading us to conclude that the present configurations are not viable candidates for the observed oscillations.

Here we study Y$_{1-x}$Ca$_x$Ba$_2$Cu$_3$O$_6$ at hole doping h=x/2=0.125. For this system 
hole injection only involves CuO$_2$ planes, without the complicacies \cite{ff4} due to oxygen doping. Calculations are carried out with a plane waves basis and ultrasoft \cite{uspp} pseudopotentials (energy cutoff 30 Ryd, 12$\times$12$\times$12 special k-point grids for 
density of states calculations, 11$\times$11$\times$11 uniform grid for Fermi surface calculations). Ca doping is described by explicit Y 
substitution in 2$\times$2$\times$1 supercells. 

\begin{figure}
\epsfxsize=2cm
\centerline{\includegraphics[clip,width=8.5cm]{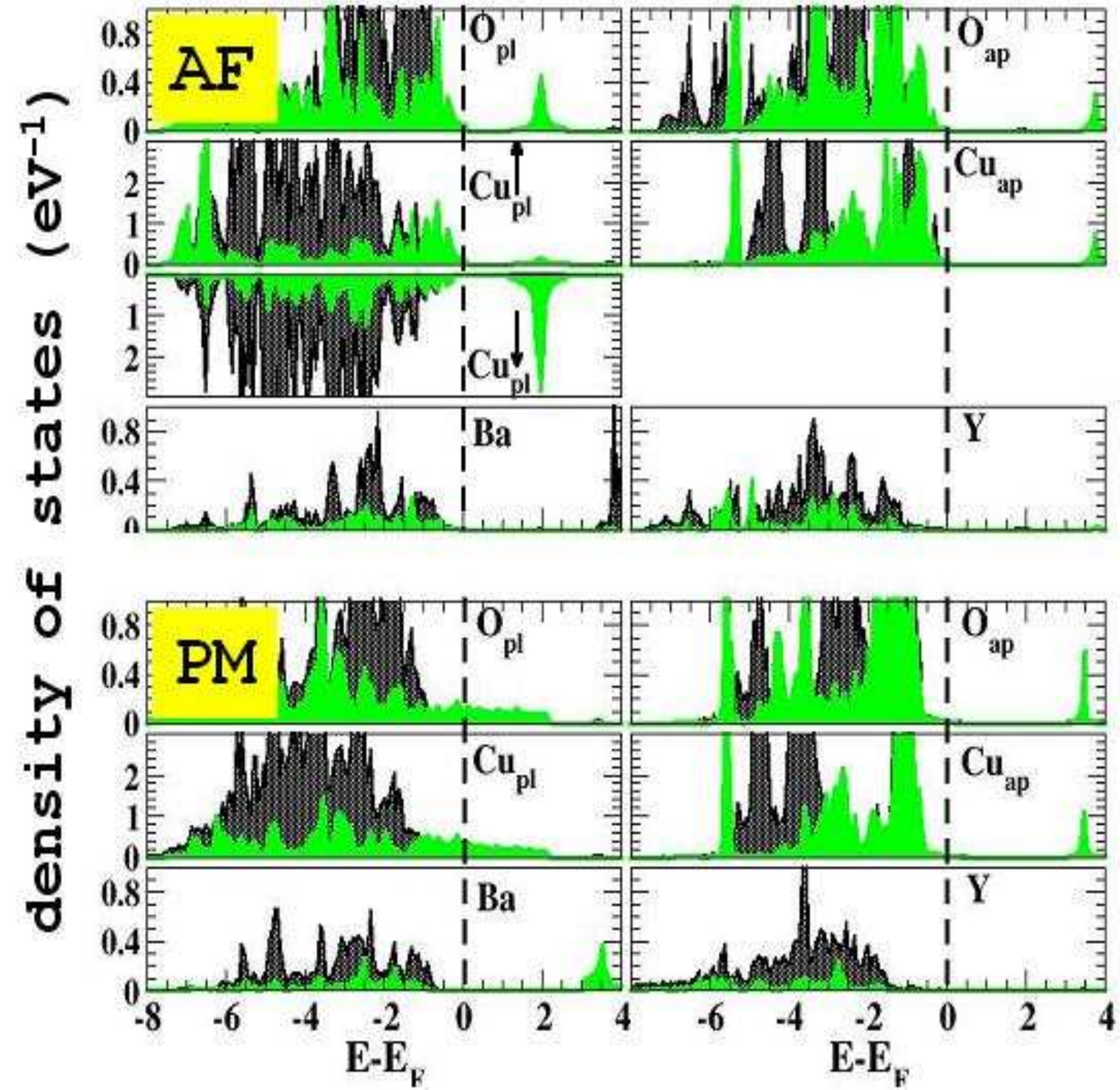}}
\caption{(Color on-line) OR-DOS for AF (top) and PM (bottom) phase at h=0.125 doping. Labels "ap" and "pl" indicate apical and 
in-plane (CuO$_2$) atoms, respectively; arrows show spin directions. For each atom the light shadowed (green) curves shows the 
most important orbital contribution (p$_x$ for O$_{pl}$, p$_z$ for O$_{ap}$, d$_{x^2-y^2}$ for Cu$_{pl}$, Ba, and Y, d$_{z^2}$ 
for Cu$_{pl}$), the dark shadowed (black) curves the contribution of the other p (for O) and d (for Cu, Y, and Ba) orbitals.
\label{dos}}
\end{figure}

The pSIC approach  successfully describes \cite{ff4}  the undoped precursor YBa$_2$Cu$_3$O$_6$ as an AF Mott insulator. The main features of the AF Mott phase can be recognized in the closely similar orbital-resolved density of states
(OR-DOS) at h=x/2=0.125 (top panels of Fig.\ref{dos}): valence and conduction bands are a mix of spin-polarized Cu d$_{x^2-y^2}$ and unpolarized 
O p$_x$p$_y$ states, with the main optical transition (involving CuO$_2$-plane orbitals) starting at 1.2 eV, and higher transitions around 
3.5 eV into apical O p$_z$ and Cu d$_{z^2}$ states. The Cu magnetic moments of 0.5 $\mu_B$ and in-plane AF coupling constant  J$\simeq$0.15 eV in  YBa$_2$Cu$_3$O$_6$ agree well with experiment. 

\begin{figure}
\epsfxsize=9cm
\centerline{\includegraphics[clip,width=8.5cm]{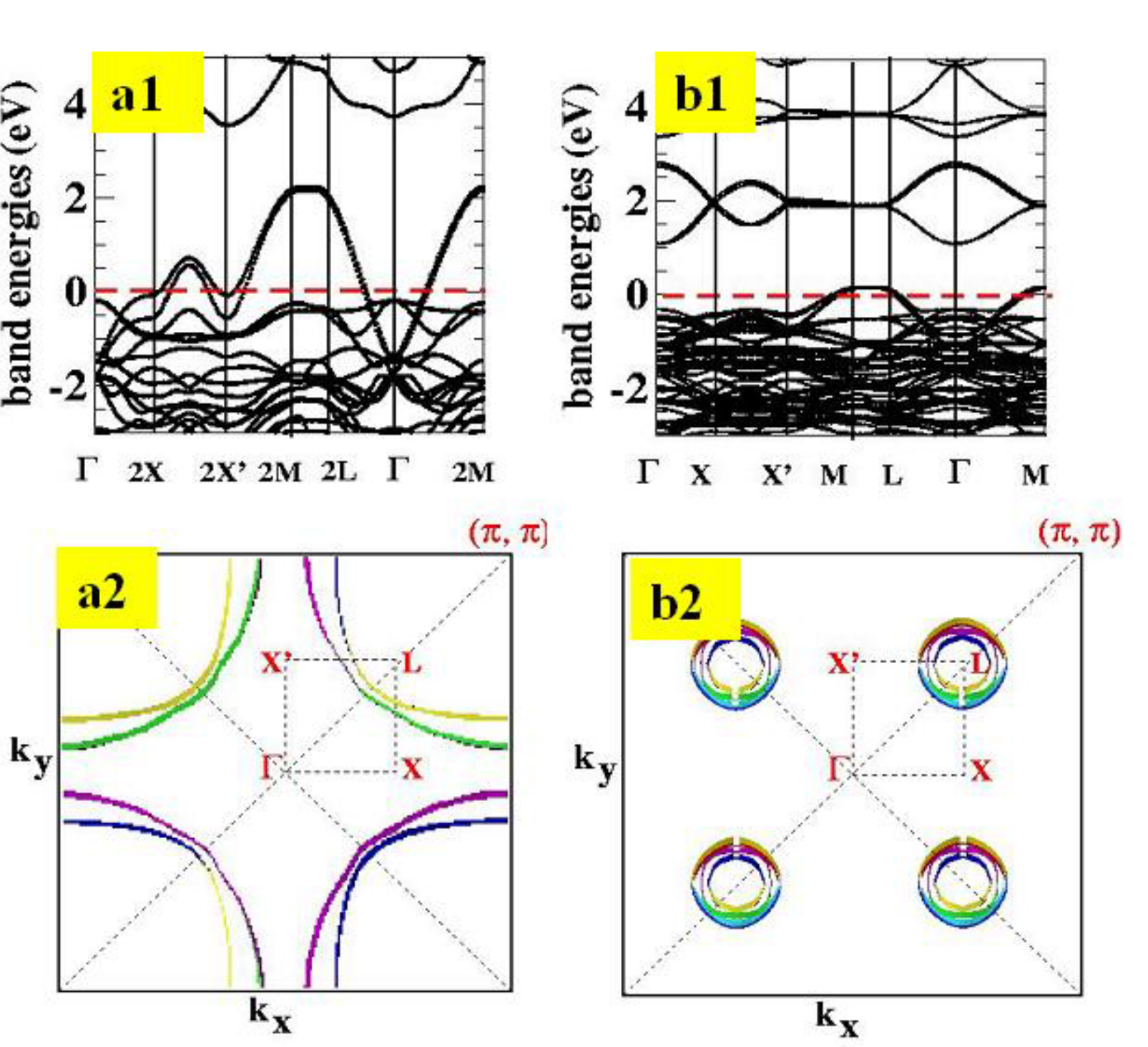}}
\caption{(Color on-line) Top: Band energies at x=1/4 (h=0.125) for (1$\times$1) PM phase (a1) and (2$\times$2) AF phase (b1). 
K-points coordinates (units of 1/a=1/b, 1/c with a, b, c unit-cell parameters) are X=[$\pi$/2,0,0], X'=[0,$\pi$/2,0], L=[$\pi$/2,$\pi$/2,0], 
M=[$\pi$/2,$\pi$/2,$\pi$/2]. Bottom: calculated FS for PM (a2) and AF (b2) phase.
\label{doped}}
\end{figure}

Figs. \ref{dos} and \ref{doped} report our calculated OR-DOS, band energies and FS for both AF and metallic non-magnetic 
(i.e. Pauli-paramagnetic (PM)) phases at h=0.125. The latter shows doubly degenerate bands, and no Mott gap. Its valence band is a 
3.5 eV-wide spin-unpolarized d$_{x^2-y^2}$-(p$_x$,p$_y$) hybrid, with Fermi level E$_{\rm F}$ at $\sim$2 eV below the valence band top (VBT). 
The corresponding FS is a large cylinder centered at the zone corner, as expected for the metallic Fermi-liquid-like state. Only in-plane atoms
contribute to the FS, whereas apical-atom states start appearing $\sim$0.2 eV below E$_{\rm F}$. In the AF state holes appear in the valence states of spin-polarized Cu, reducing their moments and weakening --though not yet destroying-- AF order. The Mott gap splits the 
valence manifold into $\sim$2 eV-wide empty upper and nearly filled lower Hubbard bands. The latter is sliced by E$_{\rm F}$ at 0.1 eV below 
its top, along the BZ diagonal near the nodal point, with off-plane states  again far below E$_{\rm F}$. AF order causes a valence bandwidth
collapse compared to the PM case, consequential to hindering nearest-neighbor Cu-Cu electron hopping. We remark that this result is  a token of self-interaction removal in our calculations, and would not be found  in a local-density-functional 
calculation \cite{ff2}.

A reason to compare the AF-phase FS with experiment is the evidence \cite{lake} that magnetic order develops in high fields in the 
superconducting state (this was modeled in Ref. \cite{zhang-field}). The FS of the AF phase consists of small-area pockets around nodal points: in fact, four circlets appear around each node, one for each of the 
two up- and two down-polarized states per 2$\times2$ CuO$_2$ plane. The average area per pocket converts to a frequency of 600 T, similar to the   SdH result in YBa$_2$Cu$_3$O$_{6.5}$ at h$\simeq$0.1. The hole-like  effective mass per pocket, however, is --0.5 m$_e$,  well below the experimental value. The carriers/holes ratio in the AF state at h=0.125 
is n/h$\sim$1.2, suggesting a weak violation of   Luttinger's sum rule (n/h=1), or more likely that the FS differs slightly from the 
Luttinger surface to which the sum rule applies \cite{dzyaloshinskii}.  Experimentally  \cite{doiron} a similar value was deduced, n/h$\sim$1.5.

\begin{figure}
\epsfxsize=9cm
\centerline{\includegraphics[clip,width=8.9cm]{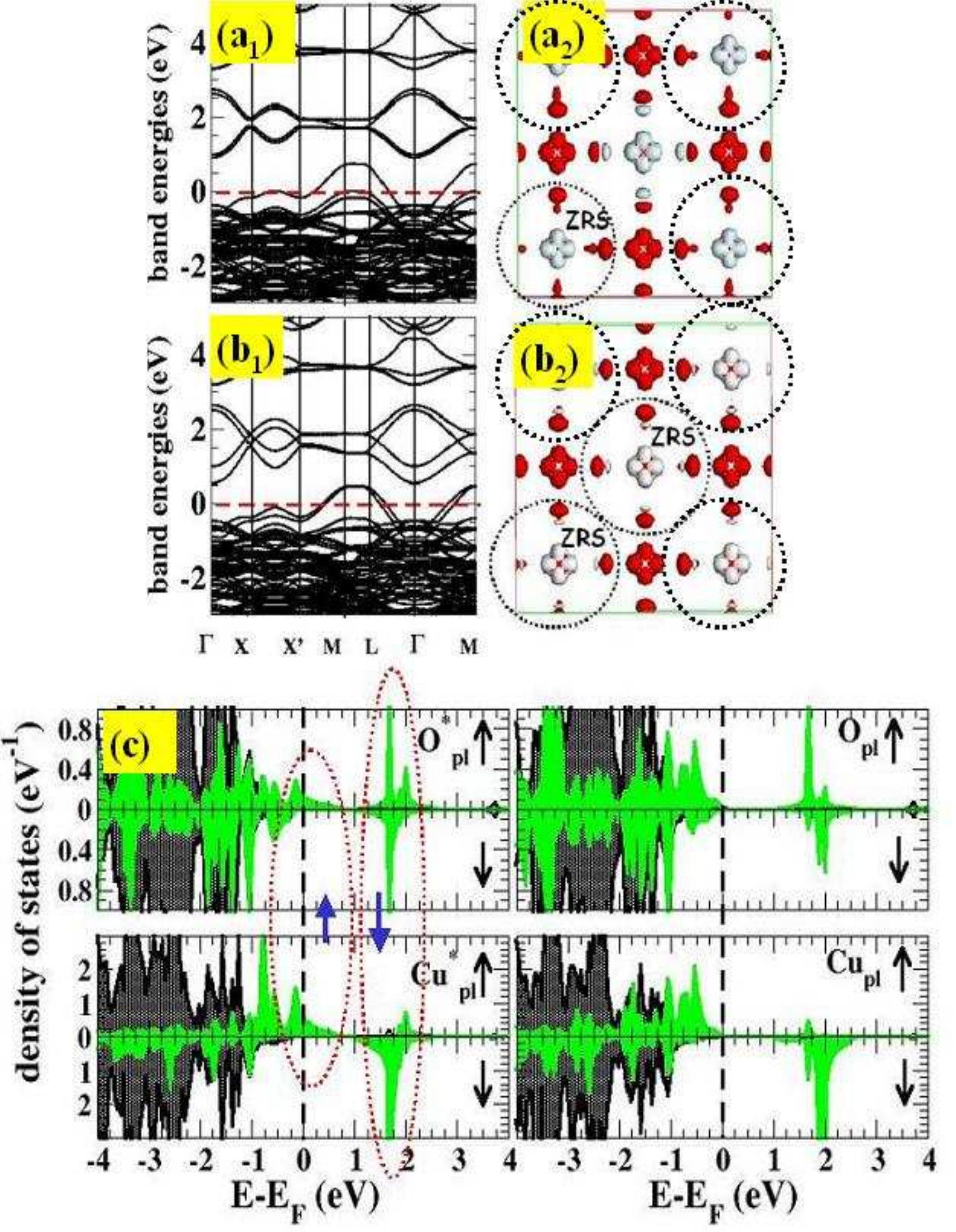}}
\caption{(Color on-line) Top: Band energies and hole spin density isosurface (h/V=0.006 bohr$^{-3}$) at h=0.125 doping for two different 
magnetic phases with ZRS included, both within (2$\times$2)symmetry. (a1) and (a2) refer to the one-ZRS per 2$\times$2 plane configuration,
(b1) and (b2) for two-ZRS per 2$\times$2 plane configuration. (c) Orbital-resolved DOS for the one-ZRS configuration  
(only planar Cu d$_{x^2-y^2}$-O (p$_x$,p$_y$) state are shown for simplicity). Red (dotted) ellipses enclose the ZRS DOS contributions from
injected and native antialigned holes (blue arrows) to give S=0 locally.
\label{zrs}}
\end{figure}

\begin{figure}[h]
\epsfxsize=9cm
\centerline{\includegraphics[clip,width=8.5cm]{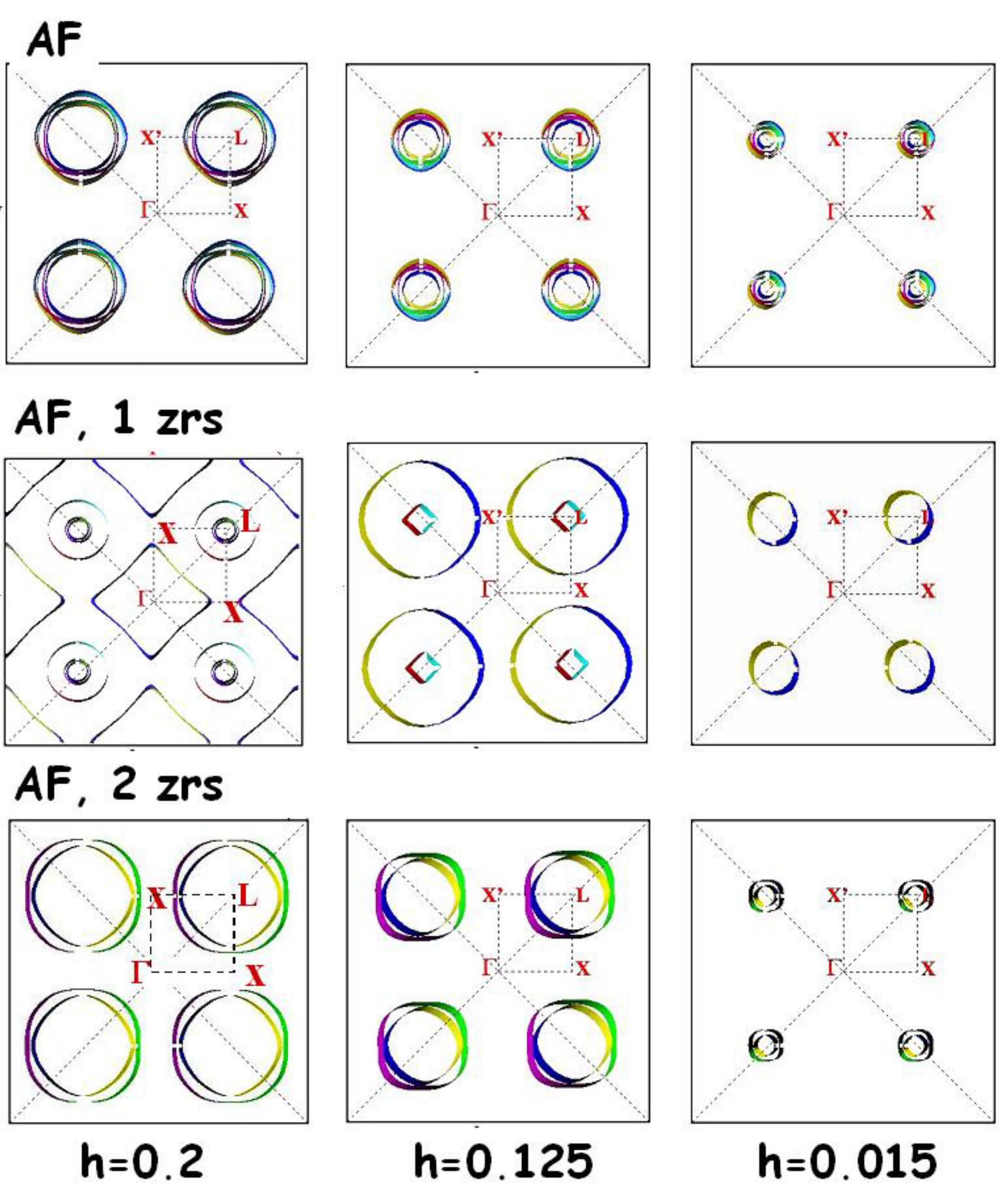}}
\caption{(Color on-line) Evolution of the FS vs. rigid Fermi level shift $\Delta$E$_{\rm F}$ for AF (top panels), AF plus one-ZRS (middle panels),
AF plus two-ZRS (bottom panels) configurations. From left to right, $\Delta$E$_{\rm F}$=--0.2 eV (h$\sim$0.2), 0 (h=0.125), and 
+0.1 eV (for AF), +0.5 eV for the two ZRS configurations (h$\sim$0.02).
\label{zrs_fs}}
\end{figure}

Next, we take a first step towards modeling  polaron states and the attendant FS, motivated by recent studies \cite{keller} on strongly lattice-coupled  polarons (a key concept in the original search for superconductivity in cuprates). In Fig.\ref{zrs_fs} (again at h=0.125 doping) we explore singlet polarons  of the Zhang-Rice type (ZRS) \cite{zhang} embedded in the CuO$_2$-plane AF background. The nature of these states is apparent from hole spin densities: in Fig.3(a2) we have one ZRS per cell (dashed circle), in Fig.3(b2) two ZRS per cell  aligned into [110]-oriented stripes. AF symmetry imposes oxygen demagnetization and   
prevents ZRS formation, so we freeze-in an O breathing distortion towards Cu in each ZRS-to-be CuO$_4$ unit (notice that the ZRS formation  does not occur without distortion).

We observe that the ZRS "condenses" on the distorted CuO$_4$ plaquette: holes localize on oxygens, and the induced magnetization is indeed 
anti-aligned to that of  Cu, producing a vanishing total magnetization on the distorted CuO$_4$ unit, as in the Zhang-Rice model. 
The ZRS signature in Fig.3(a1) is a single CuO$_2$ band being depleted (i.e. hole-filled) and lifted by about 0.7 eV over the valence bands related 
to the AF-ordered units. In the stripe configuration of Fig.3(b1), two ZRS bands are now raised by roughly the same amount above the valence AF 
background. All other bands related to non-ZRS CuO$_2$ units are well below E$_{\rm F}$ and unaffected by hole injection. Thus, in both cases, 
only the ZRS bands contribute to the FS. The OR-DOS in Fig.3(c), left panel, shows that each ZRS band involves all the four in-plane oxygens 
first-neighbors to the central Cu of the ZRS unit. This up-polarized hole couples to the native Cu down-polarized hole (the two are enclosed in 
dashed ellipses) to form the singlet. The DOS from non-ZRS CuO$_2$ units in (Fig.3(c), right panel, is close to that of the AF phase
in Fig.\ref{dos}, with hardly any magnetic moment reduction on Cu.

Fig.\ref{zrs_fs} shows  the FS of the three examined magnetic states (AF, one-ZRS, two-ZRS)  at different values of h. The calculated FS at h=0.125 is in the central-column panels (the others are discussed below).  Closed FS structures near nodal points appear to be  a general feature of spin-polarized states. The one-ZRS FS shows a large nodal-point centered pocket due to the ZRS band (plus a minor AF-background pocket). In the two-ZRS configuration we find intersecting pockets from the two ZRS bands. The pocket shape is circular for the single ZRS, while for the stripe it is the intersection of two 90$^{\circ}$-rotated ellipses. From the calculated FS areas of the ZRS bands, we obtain n/h$\sim$1.2 and $\sim$1 respectively, again mildly deviating from the Luttinger sum rule. (In  the PM and AF states,  holes are evenly distributed on all CuO$_2$ units. In the one- and two-ZRS configurations, they concentrate on one and two CuO$_2$ units respectively, so effectively doping h=0.5 and h=0.25.) The frequency for the one (two) ZRS case is $\sim$2500 (1600) T, with masses of --0.4 and --1 respectively. These values  are not to close to experiment, and the mass sign is again hole-like. 

A  definite  conclusion we can draw is that different magnetic configurations will produce FS differing in the details, but sharing the character of nodal hole pockets. This brings us again to the fact that our calculated FS pockets are hole-like, which is consistent with ARPES, but not with SdH/dHvA. To verify if electron-like branches may appear in the vicinity of E$_{\rm F}$ as calculated for h=0.125, in Fig.\ref{zrs_fs} we also plot the FS obtained rigidly shifting  E$_{\rm F}$, which roughly simulates doping fluctuations \cite{noi}. The FS remains hole-like and the pocket area 
falls sharply for an upwards shift ("decreasing doping", right panel column in Fig.4). Electron pockets at the anti-nodes do instead appear (while hole pockets open up) for E$_{\rm F}$ shifting down (left panel column in Fig.4). The shift, however, would correspond to optimal doping, i.e. to a unrealistic 50 \%  doping fluctuation. There is no obvious way to reconcile the Hall resistance sign with the present results. If one accepts the notion that the low T-high field Hall resistance (i.e. mass sign) change is a genuine electron current signature in the normal state, all the present configurations should be ruled out. 

We finally  touch upon attempts to reconcile SdH pockets --whose shape, number, and location is undetermined-- with ARPES arcs near nodes 
(see e.g.\cite{harrison,russi,norman2} for a summary). Recently, a photoemission pattern  switching from large-doping Fermi cylinders to low-doping arcs 
was observed \cite{hossain} upon tuning the self-doping of YBa$_2$Cu$_3$O$_{6+x}$ via surface treatment. Cylinders and arcs appear to coexist in different sample portions depending on the local self-doping, and arcs were hypothesized \cite{hossain,harrison} to be residual pockets 
(the estimated  "virtual" areas  would be, as our own are, in fair agreement  with SdH data). From total energies, we find the PM phase slightly lower than, but still quite close to AF. A coexistence may be expected, as suggested by $\mu$SR  experiments \cite{niedermayer}. One may then speculate that the observed FS results from a configuration-averaged superposition of pockets and cylinder portions whose shape and spectral intensity depends on extension and shape of  coexisting AF and PM regions. (Indeed, short-range AF order \cite{harrison,russi} has been shown to cause a similar  "unfocusing" of the pocket shape into an apparent arc shape.) 

In summary, we studied in detail the electronic structure of underdoped Y$_{1-x}$Ca$_x$Ba$_2$Cu$_3$O$_6$ by an innovative first-principles 
band theory givin access to  doped Mott magnetic states. AF phases (with or without embedded ZRS) produce  small closed FS branches, while Fermi-liquid cylinders are related to spinless states. At low doping the spin-polarized holes are pervasive, and the FS is composed of pockets. The pocket/arc dichotomy may be due as a superposition of pockets and cylinder portions related to competing AF and PM regions. Some features of the pockets are consistent with oscillation experiments, but at least as many are not, so that no unambiguous identification can be made. We conclude that the doped antiferromagnet and the high-density ZRS polaron configurations explored here are not those giving rise to the pockets inferred from experiments. It remains possible that more complex polaronic structures \cite{keller} be suitable candidates. 

Work supported by MiUR through PRIN 2005 and PON-Cybersar, and Fondazione Banco di Sardegna.  Calculations done on CASPUR and Cybersar clusters.



\begin{thebibliography}{99}
\bibitem{nota}
We note that it is not obvious that SdH/dHvA and ARPES  (done at high field/low T and  zero field/high T, respectively) observe the same phase. In this paper, though, we do assume in Ockham's razor fashion that one single normal (i.e. non superconducting) phase should explain both sets of measurements.
\end{thebibliography}
\end{document}